\documentclass{ws-procs975x65}  
 \begin{document}
 \newcommand{\be}{\begin{equation}}
\newcommand{\ee}{\end{equation}}
\newcommand{\parp}{\partial^+}
\newcommand{\nbar}[1]{\overline{#1}}    
\title{Maximally Supersymmetric Yang-Mills Theory\\The Story of $N=4 $ Yang-Mills Theory
}

\author{Lars Brink}

\address{Department of Fundamental Physics, Chalmers University of Technology,\\
S-412 96 G\"oteborg, Sweden\\
$^*$E-mail: lars.brink@chalmers.se}

\keywords{Style file; \LaTeX; Proceedings; World Scientific Publishing.}

\bodymatter
\section{History}
All roads lead to Rome. This historic saying is also true in fundamental physics where the Milliarium Aureum is the Yang-Mills Theory. This was not at all obvious fifty years ago. The particle physics of the 1960's was dominated by the efforts to find the theory for the strong interactions. The domineering figure was Murray Gell-Mann who also got his Nobel Prize in 1969 for "for his contributions and discoveries concerning the classification of elementary particles and their interactions". He introduced  the $SU(3)$ flavour symmetry in 1961 to classify the strongly interacting particles \cite{MGM!} and and a few years later the quarks\cite{MGM2}.

He was also one of the driving forces to understand the weak interactions among all these new particles and introduced the current algebras to study these relations \cite{GellMann:1964nh}. All these works were precursors to the Standard Model that grew out of the attemts during the 1960's and early 1970's. His first comments in this direction was in his famous $V-A$ paper from 1957\cite{Feynman:1958ty}. These attempts were however overshadowed by other attempts to find viable models in particle physics. That history is described in the scientific background text to the Nobel Prizes in physics in 2004, 2008 and 2013 \cite{2013}.

The ideas to build a field theory model for the strong interactions had failed rather miserably,  and many argued that quantum field theory was dead and that we needed something different. One major effort that was very popular in order to find a model for the strong interactions was the  {\it S-Matrix approach}. It is described in two classical text books \cite{S-matrix}. It used many consequences from quantum field theories such as crossing symmetry, analyticity, Poincar\'e invariance intertwined with new concepts such Regge asymptotic behaviour \cite{Regge:1959mz}. By introducing the elementary particles as poles in the scattering matrix it was hoped that by self-consistency one should be led to a unique S-matrix. In this approach gauge invariance  played no r\^ole and the key ingredient was analyticity rather than gauge invariance. Another one was the (seemingly)  linear Regge-trajectories of the strongly interacting particles, when the spin was plotted against the $Õm^2$ of the resonance particles with the same quantum numbers as the lowest lying state such as the $\pi$ or the $\rho$. 

In the next sections I will give a brief and by no means complete history of {\it Dual Models} and {\it String Theory} and how eventually all these attempts to construct an S-Matrix Theory led us back to the center of fundamental physics, the Yang-Mills Theories. There were  more than 1000 research papers on Dual Models in the period 1968-72\footnote{All the papers were collected by Paul Frampton}  which would justify a full review of only that field. Here I will have gauge invariance as a red thread.

\section{Dual Models}
In 1967 a new concept was introduced, "Finite Energy Sum Rules" (FESR)\cite{Dolen:1967jr}, where the authors found that a description of $\pi \,\,N$ scattering in terms of all the resonances in the direct s-channel was dual to an asymptotic description in the exchange t-channel in the sense that one should use either description but not the sum of the two. This started a major effort to find models that reproduced this {\it duality} and in 1968 Gabriele Veneziano\cite{Veneziano:1968yb} found a formula for the scattering amplitude for the process $\pi\,\pi \rightarrow \pi \,\eta$ as
\begin{align}
\nonumber A(s,t, u) = [&B(1-\alpha(t), (1-\alpha(s))\\
\nonumber +&B(1-\alpha(t), 1-\alpha(u))\\
+&B(1-\alpha(s), 1-\alpha(u))],
\end{align}
where $\alpha(s) = \frac{1}{2} + \alpha' s$, the $\rho$-trajectory, and $\alpha'$ is the slope of the trajectory, the strength of the strong interactions. $s= (p_1+p_2)^2$, $t= (p_2+p_3)^2$ and $u= (p_1+p_3)^2$ are the Mandelstam variables with $p_i$ the momenta of the four particles in the amplitude and $B$ is the Jacobi $\beta$-function.

This amplitude was an explicit solution for an S-matrix element for the process involving four particles. Could it be extended to an arbitrary number of external particles? A race all over the particle physics community was started and within less than half a year a solution was found first by Chan and Tsou \cite{Chan:1969ex}. Slightly after another way to formulate the amplitudes was found by Koba and Nielsen \cite{Koba:1969rw}. This form was very useful for the future developments that I will describe and here I give their solution

\be\label{kn}
B_N = \int_{-\infty}^{\infty} \frac{\prod_{1}{}^N \,dz_i\, \theta(z_i-z_{i+1})}{dV_{abc}}\prod_{i=1}^N \, (z_i-z_{i+1})^{\alpha_0-1}\prod_{j>i}\, (z_i-z_j)^{2 \alpha' p_i\cdot p_j}
\ee

This is a remarkable formula for the scattering of $N$ scalar particles. It has the correct resonance particles in every channel and the correct asymptotic behaviour in terms of Regge behaviour. One could also introduce isospin with the help of so-called Chan-Paton factors \cite{Paton:1969je}. In a field theoretic language one would say that this was the Born-term, the lowest order in a perturbation series but instead it was referred to as the "narrow resonance approximation". In fact there were leading scientists that claimed that one should not unitarize like in a field theory and that there must be another way of doing it. We were very far from Yang-Mills Theories!

A first problem to solve was whether the amplitudes could factorize. Indeed they could as Nambu\cite{Nambu:1969se} and Fubini, Gordon and Veneziano\cite{Fubini:1969wp} discovered. They found that the amplitudes could be written as
\be \label{factor}
B_N = <0| V\,D\,V\,......D\,V|0>
\ee
with $V$ being a three-point vertex and $D$ a propagator. The state space was now an infinite Hilbert space constructed from an infinite set of harmonic oscillators and Nambu considered that in his paper and wrote "Eq (17) suggests that the internal energy of a meson is analogous to that of a quantized string of finite length." This was the first understanding that the "Dual Model" amplitudes were indeed scattering of states of a relativistic string. Similar ideas was also put forward by Susskind \cite{Susskind:1970xm} and Nielsen\cite{holger} but it took quite some time before the Dual Models became String Theory.

Even though some of the senior physicists involved in Dual Models refused to consider them in a field theoretic framwork the form of (\ref{factor}) opened for questions about the norms of the ingoing states and about the possible symmetries of the amplitudes. The harmonic oscillators carry a spacetime index and thus there is an infinity of negative-norm states in the spectrum. Could it be that they decouple? In order for that to happen one needs an infinite symmetry in the amplitudes. Such a one was found by Virasoro\cite{Virasoro:1969zu}. However, there was a price to pay. It was only by shifting the intercept of the leading trajectory such that the lowest lying particle, the alleged $\rho$-particle is massless. A massless vector particle should ring a bell but most people involved was so impressed by the phenomenological success of the Veneziano Model that the immediate reaction was to try to move the intercept back to $1/2$. I was certainly one of those. The Virasoro algebra with the c-number discovered slightly later by Weis is
\be
[L_n, L_m] = (n-m) L_{n+m}+ \frac{d}{12}(n^3-n) \delta_{m+n,0},
\ee
where $d$ is the dimension of spacetime {which was taken for granted to be $4$, and the physical state conditions could be seen to be
\begin{align}\label{crit}
\nonumber&L_n \,| phys> = 0, \,\,n>0 \\
&(L_0 - 1)\, | phys>= 0.
\end{align}

The big issue now was if the conditions (\ref{crit}) were enough to warrant a positive-norm physical spectrum. This was solved by Brower\cite{Brower:1972wj} and Goddard and Thorn\cite{Goddard:1972iy} in 1972. They found indeed that for $d\leq 26$ the spectrum is positive definite and for the case $d=26$ the spectrum consists of only transverse states.

The first one to find that $d=26$ was special was Lovelace a year earlier\cite{Lovelace:1971fa}. He had studied possible loop diagram by just sewing tree diagrams together, which did not allow him to get the correct measure, but the correct pole structure (although with negative-norm states propagating in the loop),  and found that for a certain type of diagrams new non-unitary cuts appeared. In a brave analysis he found that if $d=26$ these discontinuities became a new set of poles with a leading trajectory with an intercept double the original one. The result was not taken too seriously then but in the light of the no-ghost theorem a year later people started to realize the importance of the "critical dimension".

The Veneziano Model only contained bosons and to have a real physical model one also needed fermions. The problem was solved by Pierre Ramond\cite{Ramond:1971gb} in a remarkable paper from the Christmas time of 1970. In the bosonic case one writes the spacetime coordinate as a function of a Koba-Nielsen variable, see (\ref{kn}), and expand it in an infinite set of harmonic oscillators. Ramond argued that the Dirac $\gamma$-matrices in a similar way should be given a dependence on a Koba-Nielsen variable and be expanded in an infinite set of anticommuting harmonic oscillators extending the Virasoro algebra into the algebra
\begin{align}
\nonumber &[L_n, L_m] = (n-m) L_{n+m}+ \frac{d}{8}(n^3-n) \delta_{m+n,0},\\
\nonumber &[L_n, F_m] = (\frac{n}{2}-m) F_{n+m}\\
&\{F_n,F_m\}=2 L_{n+m}+ \frac{d}{2}(n^2-1) \delta_{m+n,0}.
\end{align}
This is the first supersymmetry algebra, indeed it is a superconformal algebra in two dimensions in modern language. This was the key step on the road to the Superstring Theory. 

Soon after Andr\'e Neveu and John Schwarz\cite{Neveu:1971rx} constructed a new bosonic model including also anticommuting harmonic oscillators and Charles Thorn\cite{Thorn:1971jc} constructed a new model with two Ramond fermions and $N$ Neveu-Schwarz bosons. Also these models were found to be ghost-free and the critical dimension turned out to be $10$.

We now had a ghost-free model with both bosons and fermions but with massless vector particles and also a sector with massless particles with spin-2. Later they were called the open and the closed string. This was a great achievement but it was really useless for the purpose it had been invented, namely to describe strong interaction amplitudes and the interest faded. 

At this stage David Olive and myself started a programme to find out precisely if the models are unitarizable and if all amplitudes indeed are ghost-free. In order to construct correct one-loop graphs we had to follow the way that Feynman once did for Yang-Mills Theories\cite{Feynman:1963ax}. He used the so called "tree theorem" in which a loop graph can be constructed by sewing tree-amplitudes together introducing a physical-state projection operator in one propagator to ensure that no unphysical state is propagating in the loop. We constructed those physical state projection operators in all sectors\cite{Brink:1973qm} and showed that they provided another proof of the no-ghost theorem and then used these to construct all the relevant one-loop amplitudes\cite{Brink:1973gi}. They all exhibited the correct discontinuities. We also used the projection operators to show that that the couplings between open and closed strings and between bosons and fermions maintained unitarity. Even though the models still contained tachyons the models exhibited the same unitarity structure as gauge field theories.

Now it should have been  the time to take the Yang-Mills aspect seriously but it was still not the time. In fact Jo\"el Scherk had already in 1971 asked the question what happens when the slope of the Regge trajectories is taken to zero. In his first paper\cite{Scherk:1971xy} he argued that one would get a $\lambda \,\phi^3$ theory, but shortly after he found with Neveu\cite{Neveu:1971mu}  that the correct theory should be a Yang-Mills Theory. A similar question was what happens in the zero-slope limit for the closed string amplitudes? We have seen that it involves a massless spin-$2$ particle. The first one in print to connect to quantum gravity was Yoneya\cite{Yoneya:1973ca}. However slightly later Scherk and Schwarz\cite{Scherk:1974ca} put forward the idea that the open and closed strings were really extensions of a quantum theory involving gravity and Yang-Mills particles and that the intercepts were indeed correct. The idea did not really catch immediately. Still there was a hope that a new dual model would indeed bring in the strong interaction trajectories. In 1975  the next dual model was finally constructed by myself and a full football team of Italians\cite{Ademollo:1975an}. The model was even further away from strong interaction amplitudes and we interpreted the critical dimension to be $2$ hence overshooting the wanted dimension of $4$.

\section{Supersymmetric field theories and supergravity}
Ramond had found a superconformal algebra to be behind the Ramond-Neveu-Schwarz Model. A realisation of that algebra in terms of two-dimensional fields on a world-sheet was rapidly found by Gervais and Sakita\cite{Gervais:1971ji}. In 1973 Wess and Zumino asked the question if there exists a four-dimensional version of such an algebra\cite{Wess:1974tw} and found one. They also realised that one can relax the conformal invariance and only demand a Poincar\'e invariance and constructed then supersymmetric field theories. Such ones had previously been constructed in the Soviet Union by Golfand and Likhtman\cite{Golfand:1971iw} in an attempt to describe neutrinos, but that paper had not been noticed in the West. However after the papers by Wess and Zumino the field of supersymmetric four-dimensional field theory exploded. Rather quickly a supersymmetric version of Yang-Mills Theory was constructed by Ferrara and Zumino\cite{Ferrara:1974pu}. A year later Fayet\cite{Fayet:1975yi} constructed an extended $N=2$ Yang-Mills Theory.

Also Gell-Mann got interested in the problem and very quickly he classified all the superPoincar\'e algebras with Ne'man\cite{MGMN}. They particularly pointed out the CPT-invariant representations $N=4$ and $N=8$.  

These developments were quite independent from developments in string theory. This changed in 1976 when supergravity was constructed by Freedman, van Nieuwenhuzen and Ferrara\cite{Freedman:1976xh} and Deser and Zumino\cite{Deser:1976eh}. For me it was a revelation and we could use this technique to solve a problem I had worked on for quite some time, to get a formulation of the Lagrangian for strings and superstrings  as a $\sigma$-model\cite{Brink:1976sc}  \cite{Deser:1976rb}. Quickly supergravites with extended supersymmetry were constructed and it became clear that the zero-slope limit of the supersymmetric string theory as it was called then should be an extended supergravity coupled to an extended Yang-Mills Theory. However, most people working in supergravity were either relativists or field theorists deeply anchored in four spacetime dimensions while only a few of us came into the field from string theory. We did have an advantage since we  were used to work in higher dimensions. Higher dimensions had been discussed in the early 1900's first by Nordstr\"om and by Weyl and then by Kaluza and Klein but those attempts were completely forgotten in the 1970's. (Not even Gell-Mann knew that Kaluza was German and forced us to pronounce his name in Polish.) We string theorists regarded supergravity and superYang-Mills Theory as zero-slope limits of the ten-dimensional Superstring Theory, while all others thought of them as extensions of Einstein gravity and Yang-Mills Theory. Both groups were of course in retrospect right. However, the ability to work in higher dimensions was advantageous as I will now describe.
 
\section{Maximally Supersymmetric Yang-Mills Theories}

In the fall of 1976 I came to Caltech to work with John Schwarz. Knowing  that the zero-slope limit of the open superstring theory in $d=10$ is the maximally supersymmetric Yang-Mills theory, we realized that this theory should  be the mother of all supersymmetric Yang-Mills theories either by dimensional reduction or by that together with a truncation. Hence we set out to construct the ten-dimensional theory\cite{Brink:1976bc}. It is easy to write the action for it since it is really unique.

\be\label{d=10}
S= \int d^{10} x\, \{ - \frac{1}{4} F^a_{\mu \nu}F^{a\mu \nu} + i \bar \lambda^a \gamma \cdot D \lambda^a\}
\ee
with the usual definitions for the field strength and the covariant derivative.

In order for this to be supersymmetric the spinor must be both Majorana and Weyl which is possible in $d=10$. In fact this form works in three, four, six and ten dimensions with the spinors properly chosen. The supersymmetry transformations are easy to construct and also looks the same in all these dimensions.
\begin{align} \nonumber &\delta A^a_\mu = i \bar \epsilon \gamma_\mu \lambda^a - i \bar \lambda^a \gamma_\mu \epsilon, \\
\nonumber &\delta \lambda^a = \sigma^{\mu \nu} F^a_{\mu \nu} \epsilon,\\
&\delta \bar \lambda^a = - \bar \epsilon\sigma^{\mu \nu} F^a_{\mu \nu}.
\end{align}

The checking of the closure of these transformations is quite tricky though and one has to use a famous Fierz identity that we discovered.
\be
f^{abc} \bar \lambda^a \gamma_\mu \lambda^b (\gamma^\mu \lambda^c)_\alpha  = 0,
\ee
which is valid again in $d=3, 4, 6, 10$.

Now the road was open  to construct all possible supersymmetric Yang-Mills theories which we did. Here I only describe the maximally supersymmetric  theory in $d=4$ which is obtained by a straight dimensional reduction. We put $x^4, x^5,...,x^9 = 0$. The $16$-dimensional spinor is divided up into four Weyl spinors and some $\gamma$-gymnastics has to be performed leading to the following action (I write it as in the original paper.) The final action is then

\begin{align}
\nonumber S=&\int d^4x \, \{ - \frac{1}{4} F^a_{\mu \nu}F^{a\mu \nu} + i \bar \chi^a \gamma \cdot DL \chi^a + \frac{1}{2} D_\mu \Phi^a_{ij}\, D\mu \Phi^a_{ij} \\
\nonumber &-\frac{i}{2} g f^{abc}(\overline{\tilde \chi}^{ai}L\chi^{jb}\Phi^c_{ij} - \bar \chi^a_j R \tilde\chi^b_j \Phi^{ijc})\\
&-\frac{1}{4} f^2 f^{abc}\, g^{ade} \Phi^b_{ij}\Phi^c_{kl}\Phi^{ijd}\Phi^{kle}\}.
\end{align}

The resulting supersymmetry transformations are then
\begin{align}
&\nonumberÊ\delta A^a_\mu = i (\bar \epsilon_i \gamma_\mu L \chi^{ia} - \bar \chi^a_i \gamma_\mu L \epsilon^i),\\
& \nonumberÊ\delta \Phi^a_{ij} = i (\bar \epsilon_j R \tilde \chi^a_i - \bar \epsilon_i R \tilde \chi^a_j + \epsilon_{ijkl} {\overline{\tilde \epsilon}}^k L \chi^{al}),\\
&\nonumber \delta \chi^{ia} = \sigma_{\mu \nu} F^{\mu \nu a} L \epsilon^i - \gamma \cdot D \Phi^{ija}  R \tilde \epsilon_j + \frac{1}{2} g f^{abc} \Phi^{bik} \Phi^{c}_{kj} L \epsilon^j,\\
&\delta R \tilde \chi^a_i = \sigma_{\mu \nu} F^{\mu \nu a} R \tilde \epsilon_i + \gamma \cdot D \Phi^a_{ij} L \epsilon^j + \frac{1}{2} g f^{abc} \Phi^b_{ik} \Phi^{ckj} R \tilde \epsilon_j.
\end{align}
$R$ and $L$ stands for the right-handed and left-handed projections respectively and for the rest of the notation please see the original paper. 

This is the "$N=4$ theory" with an $SU(4)$ symmetry. If we had used Majorana spinors it would have been an $SO(4)$ symmetry.  Note that it is quite hard to find the four-dimensional action from scratch. We were very much helped by our knowledge of higher dimensions. However, in most circles at this time it was not appropriate to talk about higher dimensions,  so we were very careful to say that this was just a means to find the four-dimensional action.

Similar ideas were pursued at the same time by Gliozzi, Scherk and Olive\cite{Gliozzi:1976qd} who concentrated on the ten-dimensional string to make it consistent and as a by-product they also got the supersymmetric Yang-Mills theories of above. When comparing notes with Jo\"el Scherk we realized that we had the same models and hence we wrote our paper together with him.

When supersymmetric theories came  around it was noticed that the quantum properties got improved. The issue rose if such a theory could be perturbatively finite but it looked implausible.

After having constructed the $N=4$ theory we turned to other problems but in the summer of 1977 Murray Gell-Mann heard that the $\beta$-function for this theory is zero. He then prophetically declared that it is probably zero to all orders. He did not commit himself to write it in any report keeping up his promise to himself never to print anything that could be wrong. However, his comments were taken ad notam and a few months later Poggio and Pendleton\cite{Poggio:1977ma} could report that the two-loop contribution to the $\beta$-function is indeed zero. Now there was a race to compute the three-loop contribution and three years later Caswell and Zanon\cite{Caswell:1980ru} and Grisaru, Rocek and Siegel\cite{Grisaru:1980nk} could indeed confirm that it is zero. These were formidable calculations in Feynman diagrams. In the first case they had to consider some 600 diagrams while in the second case they worked with super-Feynman diagrams and had to consider 53 such diagrams. It became then clear that other techniques were needed for a general proof.

\section{The light-cone gauge formulation of $N=4$ Yang-Mills Theory}
Around 1980 and the years after I was busy with my colleagues Michael Green and John Schwarz to set up the Superstring Theory and to check its physical properties. We did it mostly in a light-cone formulation, i.e. we were only using the dynamical degrees of freedom  in a non-covariant way. In 1981 we asked what happens when we take the zero-slope limit of the one-loop graphs that we had constructed. Both for the closed string that leads to maximal supergravity and for the open string where we could isolate the maximally supersymmetric Yang-Mills Theory we could check the one-loop graph in various dimensions and see how finiteness could come about. The results were remarkably simple. Both in the supergravity case and the Yang-Mills case the complete one-loop graph for a four-particle scattering is just the box diagram of a $\phi^3$ theory with kinematical factors taking care of the spin of the particles appearing in the loop graph.

This gave me the idea to check the light-cone gauge formulation of the $N=4$ theory more explicitly and I did it with my collaborators Olof Lindgren and Bengt Nilsson\cite{Brink:1982pd}. Since the formalism we used is still not too well known I will describe the paper rather carefully. We did it by starting with the action (\ref{d=10}). We then chose the gauge $A^+=\frac{1}{\sqrt 2}\,(\,{A^0}\,+\,{A^3}\,)=0$ and solved for the kinematical field $A^-=\frac{1}{\sqrt 2}\,(\,{A^0}\,-\,{A^3}\,)$ leaving us with only the the transverse degrees of freedom. (We used $x^+$, see below, as the evolution coordinate.) Similarly for the spinor field we used the decomposition
\be 
\lambda = \frac{1}{2} (\gamma_+\gamma_- + \gamma_-\gamma_+) \lambda = \lambda_+ + \lambda_-
\ee
where again we could  solve for $\lambda_-$ since again it satisfies a kinematical equation of motion. (We did it in a path integral formulation and integrated out the non-dynamical fields). We could then rewrite the action and dimensionally reduce it to $d=4$. Finally we could introduce a superspace and rewrite the action in that space. An alternative way which we have used in many cases\cite{Bengtsson:1983pg} later is to simply look for a representation of the superPoincar\'e algebra. The final result is as follows which I now describe.

With the space-time metric $(-,+,+,\dots,+)$, the light-cone coordinates  and their derivatives are 

 \begin{align}
{x^{\pm}}=&\frac{1}{\sqrt 2}\,(\,{x^0}\,{\pm}\,{x^3}\,)\ ;\qquad ~ {\partial^{\pm}}=\frac{1}{\sqrt 2}\,(\,-\,{\partial_0}\,{\pm}\,{\partial_3}\,)\ ; \\
x =&\frac{1}{\sqrt 2}\,(\,{x_1}\,+\,i\,{x_2}\,)\ ;\qquad  {\bar\partial} =\frac{1}{\sqrt 2}\,(\,{\partial_1}\,-\,i\,{\partial_2}\,)\ ; \\
{\bar x}=&\frac{1}{\sqrt 2}\,(\,{x_1}\,-\,i\,{x_2}\,)\ ;\qquad  {\partial} =\frac{1}{\sqrt 2}\,(\,{\partial_1}\,+\,i\,{\partial_2}\,)\ ,
\end{align}
so that 
\be
{\parp}\,{x^-}={\partial^-}\,{x^+}\,=\,-\,1\ ;\qquad {\bar \partial}\,x\,=\,{\partial}\,{\bar x}\,=+1 \ .
\ee
In four dimensions,  any massless particle can be described by a  complex  field, and its complex conjugate of opposite helicity, 
the $SO(2)$ coming from the little group decomposition

\be
SO(8)\supset~SO(2)~\times~SO(6)\ .
\ee
Particles with no helicity are described by   real  fields.  The eight vector fields in ten dimension reduce to

\be
{\bf 8}^{}_v~=~{\bf 6}^{}_0+{\bf 1}^{}_1+{\bf 1}^{}_{-1}\ ,
\ee
 and the eight spinors  to

\be
{\bf 8}^{}_s~=~{\bf 4}^{}_{1/2}+{\bf \bar4}^{}_{-1/2}\ .
\ee
The representations  on the right-hand side belong to $SO(6)\sim SU(4)$, with subscripts denoting the helicity: there are six scalar fields,  
 two vector fields, four spinor fields  and their conjugates. To describe them in a compact notation, we  introduce anticommuting Grassmann variables $\theta^m$ and $\bar\theta_m$, 

\be
\{\,{\theta^m}\,,\,{{ \theta}^n}\,\}\,~=~\,\{{{\bar \theta}_m}\,,\,{\bar\theta_n}\,\}\,~=~\,\{{{\bar \theta}_m}\,,\,{\theta^n}\,\}\,~=~0\ ,
\ee
which transform as the spinor representations of $SO(6)\sim SU(4)$,

\be
\theta^m_{}~\sim~ {\bf 4}^{}_{1/2}\ ;\qquad \overline\theta_{}^m~\sim~ {\bf \bar 4}^{}_{-1/2}\ ,
\ee
where $m,n,p,q,\dots =1,2,3,4$, denote $SU(4)$ spinor indices. Their  derivatives are written as

 {\it All} the physical degrees of freedom  can be captured  in one complex superfield 

\begin{align}
\phi\,(y)=&\frac{1}{ \partial^+}\,A\,(y)\,+\,\frac{i}{\sqrt 2}\,{\theta_{}^m}\,{\theta_{}^n}\,{\nbar C^{}_{mn}}\,(y)\,+\,\frac{1}{12}\,{\theta_{}^m}\,{\theta_{}^n}\,{\theta_{}^p}\,{\theta_{}^q}\,{\epsilon_{mnpq}}\,{\partial^+}\,{\bar A}\,(y)\cr
 &~~~ +~\frac{i}{\partial^+}\,\theta^m_{}\,\bar\chi^{}_m(y)+\frac{\sqrt 2}{6}\theta^m_{}\,\theta^n_{}\,\theta^p_{}\,\epsilon^{}_{mnpq}\,\chi^q_{}(y) \ .
\end{align}
In this notation, the eight original gauge fields $A_i\ ,i=1,\dots,8$ appear as

\be
A~=~\frac{1}{\sqrt 2}\,(A^{}_1+i\,A^{}_2)\ ,\qquad \bar A~=~\frac{1}{\sqrt 2}\,(A^{}_1-i\,A^{}_2) \ ,
\ee
while the six scalar fields are written as antisymmetric $SU(4)$ bi-spinors 

\begin{align}
C_{}^{m\,4}~=~\frac{1}{\sqrt 2}\,({A^{}_{m+3}}\,+\,i\,{A^{}_{m+6}})\ ,\qquad \nbar C_{}^{m\,4}~=~\frac{1}{\sqrt 2}\,({A^{}_{m+3}}\,-\,i\,{A^{}_{m+6}})\ ,
\end{align}
for $m\;\neq\,4$; complex conjugation is akin to duality, 
\be
{{\nbar C}^{}_{mn}}~=~\,\frac{1}{2}\,{\epsilon^{}_{mnpq}}\,{C_{}^{pq}} \ .
\ee
The fermion fields are denoted by $\chi^m$ and $\bar\chi_m$. All have adjoint indices (not shown here), and are local fields in the  modified light-cone coordinates  
 \be
y~=~\,(\,x,\,{\bar x},\,{x^+},\,y^-_{}\equiv {x^-}-\,\frac{i}{\sqrt 2}\,{\theta_{}^m}\,{{\bar \theta}^{}_m}\,)\ .
\ee
This particular light-cone formulation we call $LC_2$ since  all the unphysical degrees of freedom have been integrated out, leaving only  the physical ones.

Introduce the chiral derivatives,  
\be
{d^{\,m}}=-{\partial^m}\,-\,\frac{i}{\sqrt 2}\,{\theta^m}\,{\partial^+}\ ;\qquad{{\bar d}_{\,n}}=\;\;\;{{\bar \partial}_n}\,+\,\frac{i}{\sqrt 2}\,{{\bar \theta}_n}\,{\partial^+}\ ,
\ee
which satisfy  the anticommutation relations
\be
\{\,{d^m}\,,\,{{\bar d}_n}\,\}\,=\,-i\,{\sqrt 2}\,{{\delta^m}_n}\,{\parp}\ .
\ee
One verifies  that $\phi$ and its complex conjugate $\bar\phi$ satisfy the chiral constraints
\be
{d^{\,m}}\,\phi\,=\,0\ ;\qquad {\bar d_{\,m}}\,\bar\phi\,=\,0\ ,
\ee
as well as  the ``inside-out" constraints
\be
\bar d_m^{}\,\bar d_n^{}\,\phi~=~\frac{1}{ 2}\,\epsilon_{mnpq}^{}\,d^p_{}\,d^q_{}\,\bar\phi\ ,
\ee
\be
 d^m_{}\, d^n_{}\,\bar\phi~=~\frac{1}{ 2}\,\epsilon^{mnpq}_{}\,\bar d_p^{}\,\bar d_q^{}\,\phi\ .
\ee
The Yang-Mills action is then simply
\be\int d^4x\int d^4\theta\,d^4 \bar \theta\,{\cal L}\ ,
\ee
where
\begin{align}
{\cal L}=&-\bar\phi\,\frac{\Box}{\partial^{+2}}\,\phi
~+\frac{4g}{3}\,f^{abc}_{}\,\Big(\frac{1}{\partial^+_{}}\,\bar\phi^a_{}\,\phi^b_{}\,\bar\partial\,\phi^c_{}+{\rm complex~conjugate}\Big)\cr
&-g^2f^{abc}_{}\,f^{ade}_{}\Big(\,\frac{1}{\partial^+_{}}(\phi^b\,\partial^+\phi^c)\frac{1}{\partial^+_{}}\,(\bar \phi^d_{}\,\partial^+_{}\,\bar\phi^e)+\frac{1}{2}\,\phi^b_{}\bar\phi^c\,\phi^d_{}\,\bar\phi^e\Big)\ .
\end{align}
Grassmann integration  is normalized so that $\int d^4\theta\,\theta^1\theta^2\theta^3\theta^4=1$, and $f^{abc}$ are the structure functions of the  Lie algebra.

\section{The perturbative finiteness $N=4$ Yang-Mills Theory}
After having obtained the light-cone formulation of the this theory we set out to check its perturbation expansion to see if it could be UV finite\cite{Brink:1982wv}. There are well-defined techniques to find superFeynman rules and to construct supergraphs. One direct difficulty though is that the superfield satisfies different constraints (29), (30) and (31). This means that the functional derivatives used when computing the Feynman rules become a bit intricate.
\be
\frac{\delta \phi^a(y, \theta)}{\delta \phi^b(y', \theta')}= \frac{1}{4!^2} d^4 \delta^4(x-x') \delta^4 (\theta- \theta') \delta^4(\bar \theta - \bar \theta') \delta^a_b,
\ee
\be
\frac{\delta \bar \phi^a(y, \theta)}{\delta \phi^b(y', \theta')}= 12 \frac{1}{4!^4} \frac{\bar d^4 d^4}{{\parp}^4} \delta^4(x-x') \delta^4 (\theta- \theta') \delta^4(\bar \theta - \bar \theta')\delta^a_b.
\ee
With this knowledge one can derive the expressions for the propagator, the three-point and four-point vertices and build up superFeynman diagrams. The explicit form can be seen in the paper. To estimate the naive dimension of a diagram one is helped by the fact that the $\delta$-functions and the $\theta$-integrals appearing in the propagator and in the vertex functions  are  not to be taken into account in computing the dimensionality. This fact is due to the property of supergraphs that they can always be reduced to a local expression in $\theta$. One can now check that the naive dimension of any diagram is zero. To prove finiteness one has to show that for any diagram one can integrate out some momenta. In the paper we consider a general diagram and extract either a three-point vertex or a four-point one and show that for all contributions from them one can perform this extraction. Again I refer to the paper for the details. 

The final key point is to show that one can make a Wick rotation to implement Weinberg's theorem\cite{Weinberg:1959nj}. The obstacles here are the poles in $\parp$. When we derived the formalism above we integrated out a determinant in $\parp$. This means the the remaining freedom we have is in the choice of the exact form of the poles in $\parp$. By making the choice that we interpret the pole as $(p^+ +i\epsilon p^-)^{-1}$ we indeed can make the Wick rotation and use Weinberg's theorem to complete the proof that the perturbation expansion is finite. 

At the same time as we were doing this analysis Mandelstam\cite{Mandelstam:1982cb} gave a similar proof using a slightly different light-cone formulation.

There were also other arguments put forward for finiteness around this time. Sohnius and West\cite{Sohnius:1981sn} considered anomaly multiplets, and concluded that the $N=4$ Yang-Mills theory is conformally invariant, if one can assume that the theory is supersymmetric and $O(4$)-invariant and that the structure of anomalies is given by the breakdown of conformal invariance in its coupling to supergravity. These have later been confirmed to be correct assumptions.

In the Soviet Union the group at ITEP\cite{Novikov:1983uc} attacked the problem by studying instanton calculus and could also argue that the $\beta$-function should be zero.

The proofs by us and by Mandelstam depended on a non-linear realization of supersymmetry. It was also very important to find a proof within the covariant formulation of supersymmetry. This was found by Howe, Stelle and Townsend\cite{Howe:1983sr}.

It became now an established fact that the $N=4$ theory is indeed a perturbatively finite quantum field theory. For quite some time there was a discussion if this meant that the theory is trivial. For us with a superstring background this was obviously not true since it is the zero-slope limit of the open string theory and as such is an integral part of a theory construction that was getting more and more established as the correct framework for a unified theory of all the interactions. It was amply shown by Sen\cite{Sen:1994yi} how beautiful and intricate the structure of the $N=4$ theory is when he showed the full structure of the dyons and monopoles in the theory. 

When we had proved the finiteness I went into Murray Gell-Mann's office and told him that we now had proven his conjecture. He then replied that you cannot have a field theory without a scale. This was indeed an objection to be taken seriously. In the string theory there is a scale, the slope $\alpha'$, and the Yang-Mills particles couple to other massive particles. However, the crucial observation was made by Maldacena\cite{Maldacena:1997re} when he suggested that  the $N=4$ Yang-Mills theory is dual to the superstring theory in the sense that a strong-coupling limit of one of them corresponds to the weak-coupling limit of the other. This has been very carefully studied since then and verified in all attempts. We should not think about  the $N=4$ Yang-Mills theory as just an ordinary quantum field theory with no dimensionful coupling but as a string theory in disguise and where conformal dimensions instead play an important role. Finally  the $N=4$ Yang-Mills theory had found its place in fundamental physics. A journey that started in S-Matrix theory in opposition to Yang-Mills theories gave us string theory and the Superstring Theory and generalized Yang-Mills theories to end back into the new Millarium Aureum of fundamental physics,  the $N=4$ Yang-Mills theory.

\end{document}